\documentclass[11pt,letterpaper]{article}
\pdfoutput=1
\usepackage{a4wide,epsfig,amsmath,amssymb,cite,scalefnt,graphicx}
\numberwithin{equation}{section}
\allowdisplaybreaks 
\usepackage{array}
\usepackage{epstopdf}
\usepackage{tikz}
\usetikzlibrary{intersections}
\usetikzlibrary{calc}
\usetikzlibrary{shapes}
\usetikzlibrary{arrows,shapes,positioning}
\usetikzlibrary{decorations.markings}
\tikzstyle arrowstyle=[scale=1]
\tikzstyle directed=[postaction={decorate,decoration={markings,
    mark=at position .5 with {\arrow[arrowstyle]{stealth}}}}]
\tikzstyle reverse directed=[postaction={decorate,decoration={markings,
    mark=at position .5 with {\arrowreversed[arrowstyle]{stealth};}}}]

\def\be{\begin{equation}}
\def\ee{\end{equation}}
\def\pa{\partial}
\def\Ocal{{\cal O}}

\def\nn{\nonumber\\}

\input{epsf}

\begin{document}

\begin{titlepage}
\begin{flushright}
LTH1284\\
{\today
}\\

\end{flushright}
\date{}
\vspace*{3mm}

\begin{center}
{\Huge Scaling dimensions at large charge for cubic $\phi^3$ theory in six dimensions}\\[12mm]
{\bf I.~Jack\footnote{{\tt dij@liverpool.ac.uk}} and  D.R.T.~Jones\footnote{{\tt drtj@liverpool.ac.uk}} 
}\\

\vspace{5mm}
Dept. of Mathematical Sciences,
University of Liverpool, Liverpool L69 3BX, UK\\

\end{center}

\vspace{3mm}
\begin{abstract}
The $O(N)$ model with scalar quartic interactions at its ultraviolet fixed point, and the $O(N)$ model with scalar cubic interactions at its infra-red fixed point are conjectured to be equivalent. This has been checked by comparing various features of the two models at their respective fixed points. Recently, the scaling dimensions of a family of operators of fixed charge $Q$ have been shown to match at the FPs up to $\Ocal\left(\frac{1}{N^2}\right)$at leading order (LO) and next-to-leading order (NLO) in $Q$ using a semiclassical computation which is valid to all orders in the coupling. Here we perform a complementary but overlapping comparison using a perturbative calculation in six dimensions, up to three-loop order in the coupling, to compare these critical scaling dimensions beyond NLO in $Q$, in fact to all relevant orders in $Q$. We also obtain the corresponding results at $\Ocal\left(\frac{1}{N^3}\right)$ for the cubic theory.
\end{abstract}

\vfill

\end{titlepage}

\setcounter{footnote}{0}

\section{Introduction}
There has been remarkable progress in recent years in the study of scalar field theories characterised by renormalisable self-interactions. As well as increasing precision of perturbative calculations (up to the seven-loop level\cite{che1,che2,Klein2,Kom1,Kom2,Kom3,schnetz}) the use of semi-classical approximations has proved very fruitful, both in direct comparison with perturbation theory and, importantly, in exploring areas of parameter space inaccessible to perturbation theory for the relevant theories. In Ref.~\cite{fei2} it was conjectured that the $O(N)$ model with scalar quartic interactions (the ``quartic theory'') at its ultraviolet fixed point (FP) is equivalent to the $O(N)$ model with scalar cubic interactions (the ``cubic theory'') at its infra-red FP (we shall call the theory at its FP the ``critical theory''). It was shown that the $\frac{1}{N}$ expansions of various operator scaling dimensions in the critical cubic theory match the known results for the critical quartic theory continued to $d=6-\epsilon$ dimensions. This comparison was further refined by a three-loop calculation in Ref.~\cite{fei1}. Subsequently four-loop\cite{jag1} and five-loop\cite{komp,komgra} calculations were performed for the cubic theory, and once again the critical scaling dimensions were found to agree at these orders with the known results in the $\frac{1}{N}$ expansion for the critical quartic theory.

Meanwhile, following early work in Ref.~\cite{son}, there has been considerable recent interest\cite{horm}-\cite{Bad} in the use of semiclassical methods to investigate the scaling dimensions of composite operators. This allows results to all orders in the coupling but at leading order (LO) and next-to-leading order (NLO) in the charge $Q$ of the operator. Of particular relevance for our purposes, in Ref.~\cite{rod2} the authors computed the scaling dimensions of traceless symmetric operators of charge $Q$ (which we shall denote by $T_Q$) in both the cubic and quartic critical theories for $d=6-\epsilon$ at leading order in both $Q$ and $\epsilon$, and found agreement at this level. Recently systematic semi-classical calculations of these scaling dimensions at both leading and non-leading order in $Q$ have been performed for the quartic\cite{hyman} and cubic\cite{sann4} theories, and agreement has been found in the critical theories for $d=6-\epsilon$ to high orders in $\frac{Q\epsilon}{N}$. Specifically the two $\Ocal\left[Q\left(\frac{Q\epsilon}{N}\right)^j\right]$ results were shown to agree for $j=0\ldots 8$ and the two $\Ocal\left[N\left(\frac{Q\epsilon}{N}\right)^{j+1}\right]$ results were shown to agree for $j=0\ldots 6$; we may expect that the results for higher values of $j$ will continue to agree\footnote{The scaling dimensions were also shown in Ref.~\cite{sann4} to match in a large $Q/N$ expansion.}. These two sets of terms represent leading order contributions in $\frac{1}{N}$. For low values of $j$ ($j=0,1,2$) results were also obtained in the critical cubic theory for some of the terms subleading in $N$ compared to these; though those deriving from the NLO computation had to be computed numerically. These results for $j=0,1,2$ for the critical cubic theory were compared with previous results obtained at $\Ocal\left(\frac{1}{N}\right)$\cite{step} and $\Ocal\left(\frac{1}{N^2}\right)$\cite{derk} in the $\frac{1}{N}$ expansion for the critical scaling dimensions of charged operators in the quartic theory. When these latter results were expanded in $\epsilon$ around $d=6-\epsilon$, agreement was obtained up to the limits of numerical precision wherever a comparison was possible. In the present article we perform an overlapping but complementary calculation to that of Ref.~\cite{sann4}. We compute the scaling dimension of $T_Q$ perturbatively for the cubic theory in $d=6-\epsilon$ up to three loop order in the coupling. At the critical point, this corresponds to terms up to $\Ocal(\epsilon^3)$. Organising the critical anomalous dimensions of $T_Q$ as a power series in $\frac1N$, we may compare with the $\Ocal\left(\frac{1}{N}\right)$ results of Ref.~\cite{step} and the $\Ocal\left(\frac{1}{N^2}\right)$ results of Ref.~\cite{derk}, after expanding these latter in $\epsilon$ around $d=6$ and keeping terms up to $\epsilon^3$; we find precise agreement for all relevant powers of $Q$. We also obtain the corresponding results at $\Ocal\left(\frac{1}{N^3}\right)$ for the critical cubic theory, potentially providing useful comparison with future extensions of the $\frac1N$ expansion in the quartic theory.

The paper is organised as follows: in Sect.~2 we describe the cubic and quartic $O(N)$ models and review the known results in the $\frac1N$ expansion for the scaling dimensions of the field $\phi$ and the fixed-charge operator $T_Q$ in the critical quartic theory. In Sect.~3 we describe our own calculation for the scaling dimension of $T_Q$ up to three loops in the cubic theory. We then specialise to the critical theory and compare with the corresponding results for the critical quartic theory. We relegate some details of the calculation to the Appendix. Here we list both the Feynman diagram results we have computed ourselves and for convenience those we have adopted from Ref.~\cite{fei1}. We also give results for the scaling dimensions in terms of individual diagrams, as an aid to following or checking our calculations.

\section{The quartic and cubic $O(N)$ models}
In this section we introduce both the quartic and cubic scalar $O(N)$ models whose properties we shall be comparing at their critical points.
The quartic scalar action is given by
\be
S=\int d^dx(\frac12\pa_{\mu}\phi^i\pa^{\mu}\phi^i+\frac12\sigma\phi^i\phi^i-\frac{3}{2\lambda}\sigma^2),
\label{Squart}
\ee
where the sum over $i$ runs from $1$ to $N$. This theory is renormalisable in $d=4-\epsilon$ dimensions. We shall only list here the particular properties at the critical point in which we are interested; for a more complete exposition see, for instance, Ref.~\cite{klein}. At the critical point the final term may be neglected; and the anomalous dimension of the field $\phi$ is expressed in terms of the critical index $\eta$ as
\be
\gamma_{\phi}=\frac12\eta.
\label{gamdef}
\ee 
$\eta$ may be computed in the large-$N$ expansion as
\be
\eta=\sum_i\frac{\eta_i}{N^i}.
\ee
Here $\eta_1$ is given by
\be
\eta_1=-4\frac{\Gamma(2\mu-2)}{\Gamma(2-\mu)\Gamma(\mu-1)\Gamma(\mu-2)\Gamma(\mu+1)},
\label{eta1}
\ee
where $d=2\mu$, and $\eta_2$ is given by\cite{vas1}\cite{vas2}
\be
\eta_2=\eta_1^2(T_1+T_2+T_3),
\label{eta2}
\ee
where
\begin{align}
T_1=&R_1+\frac{\mu^2+\mu-1}{2\mu(\mu-1)},\nn
T_2=&\frac{\mu}{2-\mu}R_1+\frac{\mu(3-\mu)}{2(2-\mu)^2},\nn
T_3=&\frac{\mu(2\mu-3)}{2-\mu}R_1+\frac{2\mu(\mu-1)}{2-\mu}.
\label{eta2b}
\end{align}
Here
\be
R_1=\psi(2-\mu)+\psi(2\mu-2)-\psi(2)-\psi(\mu-2),
\ee
where $\psi$ is the digamma function.

 An operator of charge $Q$ is given by
\be
T_{Q}=T_{i_1i_2\ldots i_{Q}}\phi_{i_1}\phi_{i_2}\ldots \phi_{i_{Q}},
\label{Tdef}
\ee
where $T_{i_1i_2\ldots i_{Q}}$ is symmetric,  and traceless on any pair of indices.
The scaling dimension of $T_Q$ in the quartic theory at the FP is given by
\be
\Delta_Q=\left(\frac{d}{2}-1\right)Q+\gamma_Q+Q\gamma_{\phi},
\label{deldef}
\ee
where the first term is the classical scaling dimension, $\gamma_{\phi}$ is given by Eq.~\eqref{gamdef}, and the anomalous dimension $\gamma_Q$ of $T_Q$ has been computed in the $1/N$ expansion as\cite{step,derk}
\begin{align}
\gamma_Q=&-\frac1N\frac{\mu}{2(\mu-2)}\eta_1Q(Q-1)\nn
&-\frac{1}{N^2}\eta_1^2\frac{Q(Q-1)\mu}{4(\mu-1)(\mu-2)^2}\Bigl\{2(Q-2)\mu(\mu-1)^2[\psi'(1)-\psi'(\mu)]\nn
&+\mu(2\mu-3)-2(\mu-1)(2\mu^2-3\mu+2)R_2\Bigr\}+\ldots.
\label{gamQ}
\end{align}
Here $\eta_1$ is again given by Eq.~\eqref{eta1}, and 
\be
R_2=\psi(2-\mu)+\psi(2\mu-2)-\psi(1)-\psi(\mu-1).
\ee
 The ellipsis indicates higher-order terms in the $1/N$ expansion.

The cubic action is given by
\be
S=\int d^dx(\frac12\pa_{\mu}\phi^i\pa^{\mu}\phi^i+\frac12\pa_{\mu}\sigma\pa^{\mu}\sigma+\frac12g\sigma\phi^i\phi^i+\frac16h\sigma^3),
\label{Scube}
\ee
where the sum over $i$ again runs from $1$ to $N$. This theory is renormalisable in $d=6-\epsilon$ dimensions. However, the two theories given by Eqs.~\eqref{Squart}, \eqref{Scube} are believed to be equivalent at their conformal FPs. In the next section we shall compute perturbatively the scaling dimension $\Delta_Q$ of the operator $T_Q$ in Eq.~\eqref{Tdef} at the FP of the cubic theory, in order to compare with the corresponding result for the quartic theory, given later in Eq.~\eqref{Delquar}. For this we shall need the expressions for the values of the couplings at the FP in the cubic theory, as obtained via the $\epsilon$ expansion. These are given by\cite{fei2,fei1}
\begin{align}
g_*=&\sqrt{\frac{6\epsilon}{N}}\Bigl[1+\frac{22}{N}+\frac{726}{N^2}-\frac{326180}{N^3}+\left(-\frac{155}{6N}-\frac{1705}{N^2}+\frac{912545}{N^3}\right)\epsilon\nn
&+\frac{1777}{144N}\epsilon^2+\frac{1}{N^2}\left(\frac{29093}{36}-1170\zeta_3\right)\epsilon^2+\ldots\Bigr],\nn
h_*=&6\sqrt{\frac{6\epsilon}{N}}\Bigl[1+\frac{162}{N}+\frac{68766}{N^2}+\frac{41224420}{N^3}+\left(-\frac{215}{2N}-\frac{86335}{N^2}-\frac{75722265}{N^3}\right)\epsilon\nn
&+\frac{2781}{48N}\epsilon^2+\frac{1}{6N^2}(270911-157140\zeta_3)\epsilon^2+\ldots\Bigr].
\label{fixval}
\end{align}
We shall also need expressions for the quantities from the quartic theory, expanded around $d=6-\epsilon$. 
Using Eqs.~\eqref{eta1} and \eqref{eta2}, the terms $\eta_1$, $\eta_2$ in the $1/N$ expansion of $\eta$ have the expansions
\begin{align}
\eta_1=&2\epsilon-\frac{11}{6}\epsilon^2-\frac{13}{72}\epsilon^3+\ldots,\nn
\eta_2=&88\epsilon-\frac{835}{3}\epsilon^2+\frac{6865}{36}\epsilon^3+\ldots
\label{epsex}
\end{align}

Using Eqs.~\eqref{gamdef}, \eqref{gamQ}, $\Delta_Q$ in Eq.~\eqref{deldef} may be expanded in $\epsilon$ as 
\begin{align}
\Delta_Q=&\left(2-\frac{\epsilon}{2}\right)Q\nn
&+\frac{1}{N}\left[(-3Q^2+4Q)\epsilon+\left(\frac74Q^2-\frac83Q\right)\epsilon^2+\left(\frac{11}{16}Q^2-\frac79Q\right)\epsilon^3\right]\nn
&+\frac{1}{N^2}\Bigl\{44(-3Q^2+4Q)\epsilon+\left(-45Q^3+\frac{857}{2}Q^2-\frac{1568}{3}Q\right)\epsilon^2\nn
&+\left[\left(36\zeta_3+\frac{93}{4}\right)Q^3-\left(108\zeta_3+\frac{3743}{24}\right)Q^2+\left(72\zeta_3+\frac{4105}{18}\right)Q\right]\epsilon^3\Bigr\}+\ldots
\label{Delquar}
\end{align}
which we shall compare with our perturbative results for the critical cubic theory in the next section.

\section{Loop calculations}
\begin{figure}
\centering
\includegraphics[width=0.25\columnwidth]{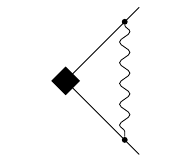}
\caption{One-loop diagram for $\gamma_{T_{Q}}$}\label{diagone}
\end{figure}

In this section we perform the perturbative computation of the scaling dimension of an operator of the form \eqref{Tdef} within the cubic theory given by Eq.~\eqref{Scube}, up to three loops, and then compare the result at the FP Eq.~\eqref{fixval} with the corresponding result for the quartic theory in Eq.~\eqref{Delquar}. The scaling dimension of $T_Q$ is given by an equation of the same form as Eq.~\eqref{deldef} as
\be
\Delta_Q=\left(\frac{d}{2}-1\right)Q+\gamma_Q+Q\gamma_{\phi},
\label{Delfull}
\ee
where $\gamma_Q$ is again the anomalous dimension of $T_Q$ and $\gamma_{\phi}$ the anomalous dimension of the single field $\phi$; but now all quantities are computed in the cubic theory. The $L$-loop one-particle-irreducible (1PI) diagrams contributing to $\gamma_Q$ are constructed from a single $T_Q$ vertex and a number of three-point vertices derived from the action in Eq.~\eqref{Scube}. The latter vertices thus have either two $\phi$ lines and one $\sigma$ line, or three $\sigma$ lines, emerging from them; which may form either internal or external lines in the diagram. These $L$-loop diagrams have up to $L+1$ internal lines emanating from the $T_Q$ vertex, together with two or three internal lines emanating from the other vertices; the total number of external lines is always the same as the number of internal lines emerging from the $T_Q$ vertex. These correspond to logarithmically divergent Feynman integrals. The Feynman diagrams with no more than three internal lines emerging from any vertex (including the $T_Q$ vertex) have the same topology as those contributing to the $\beta$-function computation, and their pole terms have already been derived up to three loops and given in Refs.~\cite{fei1}. As an example, in Fig.~\ref{diagone} we show the one-loop diagram. Straight and wavy lines denote $\phi$ and $\sigma$ propagators respectively; and moreover we use a small circle to denote the cubic vertices derived from Eq.~\eqref{Scube}, and a lozenge to denote the $T_Q$ vertex.  Fig.~\ref{diagone} has the same topology and therefore the same pole structure as diagram (a) in Fig.~8 of Appendix B in Ref.~\cite{fei1}. 
It therefore produces a contribution to $\gamma_Q^{\rm{1PI}}$given by 
\be
\gamma_Q^{(1)}=\frac12Q(Q-1)c_ag^2,
\label{Del1}
\ee
where $c_a$ is the simple pole coefficient from the diagram (a) in Fig. 8  of Ref.~\cite{fei1}. For Eq.~\eqref{Delfull} we also need the perturbative computation of the anomalous dimension of $\phi$, $\gamma_{\phi}$. The computation of the anomalous dimensions for the scalar fields $\phi$ and $\sigma$ in the cubic theory was given explicitly in Ref.~\cite{fei1}, similarly in a helpful diagram-by-diagram form. With a view to a uniform presentation, we have reconstructed the expression for $\gamma_{\phi}$ in terms of simple pole contributions (we have no need for $\gamma_{\sigma}$), so at one loop we have
\be 
\gamma_{\phi}^{(1)}=-\frac12c_Ag^2
\label{gam1}
\ee
where in our notation $c_A$ is the simple pole coefficient from the diagram (a) in Fig.~7 of Ref.~\cite{fei1}. We give the result for $\gamma_{\phi}$ in a corresponding form up to three loops in the Appendix. Combining Eqs.~\eqref{Del1}, \eqref{gam1}, we find from Eq.~\eqref{Delfull}
\be
\Delta_Q=\frac12Q(Q-1)c_ag^2-\frac12Qc_Ag^2,
\label{Del1a}
\ee
The values for the simple pole coefficients can be obtained from Ref.~\cite{fei1}; but for completeness we have also listed them in the Appendix, in Eqs.~\eqref{simpa}, \eqref{simpA}. We emphasise that by $c_a$ we mean precisely the simple pole coefficient from the Feynman integral, whereas the pole terms given in Figs.~7-9 of Ref.~\cite{fei1} also include a symmetry factor. After inserting the values of the simple pole coefficients, we obtain at one loop
\be
\Delta_Q^{(1)}=-\frac12Q(Q-1)g^2+\frac16Qg^2.
\ee
In order to save space we have refrained from depicting most of the diagrams contributing to $\Delta_Q$ at two and three loops. Specifically, we do not show  those which can straightforwardly be reconstructed from the diagrams shown in Figs.~8 and 9 of Ref.~\cite{fei1}, by considering the various ways in which a single $T_Q$ vertex (with two or three internal legs), and cubic vertices derived from Eq.~\eqref{Scube}, may be assigned to each diagram. However, at three loops there are also diagrams with four internal legs emerging from the $T_Q$ vertex, which are not considered in Ref.~\cite{fei1}, since of course such diagrams do not feature in a $\beta$-function computation for a theory with only cubic interactions. These diagrams, which furnish the leading contribution in $Q$ at three loops, are depicted in Fig.~\ref{diagtwo} in the Appendix. We have therefore been obliged to compute these Feynman diagrams ourselves; the results are likewise given in the Appendix. The results for $\Delta_Q$ in terms of the simple pole coefficients at two and three loops are given in Eqs.~\eqref{Del2}, \eqref{Del3}, with Eqs.~\eqref{a2}, \eqref{a3}; in the absence of the full set of diagrams, these detailed results should facilitate following our calculations. Upon inserting the values of these coefficients from Eqs.~\eqref{simpa}, \eqref{simpA}, \eqref{dcoeff}, we obtain 
\begin{align}
\Delta_Q^{(2)}=&-Q(Q-1)(Q-2)(\frac14g^4+\frac16g^3h)\nn
&+Q(Q-1)\Bigl[\frac{7}{144}Ng^4-\frac{49}{144}g^4-\frac14g^3h+\frac{7}{144}g^2h^2\Bigr]\nn
&+Q\left[-\frac{11}{432}Ng^4++\frac{13}{216}g^4+\frac19g^3h+\frac{11}{432}g^2h^2\right],
\label{Dela}
\end{align}
and
\begin{align}
\Delta_Q^{(3)}=&-Q(Q-1)(Q-2)(Q-3)\left(\frac14g^6+\frac14g^5h+\frac18g^4h^2\right)\nn
&+Q(Q-1)Q-2)\Bigl[Ng^6\left(\frac16\zeta_3-\frac{35}{288}\right)+\frac{11}{288}Ng^5h\nn
&-\left(\zeta_3+\frac{7}{18}\right)g^6-\frac{49}{36}g^5h-\left(\zeta_3-\frac{89}{288}\right)g^4h^2
+\frac16\left(\zeta_3-\frac{35}{48}\right)g^3h^3\Bigr]\nn
&+Q(Q-1)\Bigl[\frac{449}{2592}Ng^6+\frac{73}{432}Ng^5h-\frac{119}{5184}Ng^4h^2+\frac{11}{3456}N^2g^6\nn
&-\frac{749}{324}g^6-\left(\zeta_3-\frac{37}{432}\right)g^5h+\frac12\left(\zeta_3-\frac{8131}{2592}\right)g^4h^2+\frac{185}{864}g^3h^3+\frac{143}{10368}\Bigr]\nn
&+Q\Bigl[\frac{13}{31104}N^2g^6-\frac{29}{3888}Ng^6-\frac{49}{576}g^5h+\frac{193}{15552}g^4h^2\nn
&-\frac16\left(\zeta_3-\frac{1133}{648}\right)g^6-\frac{17}{162}g^5h-\frac16\left(\zeta_3-\frac{5881}{2592}\right)g^4h^2-\frac{157}{5184}g^3h^3-\frac{109}{10368}g^2h^4\Bigr].
\label{Delb}
\end{align}
Finally, upon inserting the FP values for $g$ and $h$ given by Eq.~\eqref{fixval}, we obtain
\begin{align}
\Delta_Q=&\left(2-\frac{\epsilon}{2}\right)Q\nn
&+\frac{1}{N}\left[(-3Q^2+4Q)\epsilon+\left(\frac74Q^2-\frac83Q\right)\epsilon^2+\left(\frac{11}{16}Q^2-\frac79Q\right)\epsilon^3+\ldots\right]\nn
&+\frac{1}{N^2}\Bigl\{44(-3Q^2+4Q)\epsilon+\left(-45Q^3+\frac{857}{2}Q^2-\frac{1568}{3}Q\right)\epsilon^2\nn
&+\left[\left(36\zeta_3+\frac{93}{4}\right)Q^3-\left(108\zeta_3+\frac{3743}{24}\right)Q^2+\left(72\zeta_3+\frac{4105}{18}\right)Q\right]\epsilon^3+\ldots\Bigr\}\nn
&+\frac{1}{N^3}\Bigl\{1936(-3Q^2+4Q)\epsilon+\left(-9000Q^3+59520Q^2-66872Q\right)\epsilon^2\nn
&+\Bigl[-1350Q^4+\left(4536\zeta_3+20574\right)Q^3-\left(3996\zeta_3+88683\right)Q^2\nn
&-\left(4212\zeta_3-\frac{193285}{2}\right)Q\Bigr]\epsilon^3+\ldots\Bigr\}+\ldots
\label{Delcub}
\end{align}
This agrees with Eq.~\eqref{Delquar} as far as terms of order $\frac{1}{N^2}$; the results at $\Ocal\left(\frac{1}{N^3}\right)$ are not yet available for the quartic theory for general values of $Q$. However the agreement at least persists for $Q=1$, since for this value the computation reduces to that of the scaling dimension for a single field $\phi$; and as explained in more detail in the Appendix, the anomalous dimension of $\phi$ was shown in Ref.~\cite{fei1} to agree up to $\Ocal\left(\frac{1}{N^3}\right)$ in the cubic and quartic theories. Moreover the leading $\frac{Q^4\epsilon^3}{N^3}$ term was already obtained from the semiclassical calculations in Refs.~\cite{hyman,sann4}.

The ellipses in Eq.~\eqref{Delcub} indicate terms originating from higher-loop contributions. Since these are eighth-order and higher in the couplings, we see from Eq.~\eqref{fixval} that they are $\Ocal(\epsilon^4)$ and higher; but they may appear at any order in $1/N$, bearing in mind that factors of $N$ may be produced by tensor contractions within diagrams, as may be seen in Eq.~\eqref{Del2}, for instance.

\section{Conclusions}

In this paper we have made a detailed comparison in $d=6-\epsilon$ dimensions between perturbation 
theory for $\phi^3$ theory and large-$N$ results for $\phi^4$ theory, at the fixed points of the respective theories. We showed that, for the scaling dimension of a set of operators of fixed charge $Q$, this comparison may be carried out to all relevant orders in $Q$ at $\Ocal\left(\frac1N\right)$ and $\Ocal\left(\frac{1}{N^2}\right)$.  The  results support the conjecture that essentially the renormalisable theories in $d=4-\epsilon$ and $d=6-\epsilon$ correspond to the same conformal theory at their respective fixed points, and provide data on the $d=6-\epsilon$ side for further checks at $\Ocal\left(\frac{1}{N^3}\right)$ when corresponding results become available for $d=4-\epsilon$. It would be interesting to explore further the relationship between these two theories at these FPs and the other simple renormalisable scalar theory, to wit $\phi^6$ in $d=3-\epsilon$. This paper and the general canon of research to which they contribute have obvious relevance to the theory of critical phenomena. A different  question is whether these techniques could be applied to other renormalisable, classically scale invariant theories such as, in particular, 
QCD. This will involve formidable technical obstacles evaded in the elementary scalar case, but clearly any progress in the understanding of QCD remains a primary goal for particle physicists. 
\section*{Acknowledgements}

 We are grateful to John Gracey for helpful comments, and the referee for useful suggestions. DRTJ thanks the Leverhulme Trust for the award
of an Emeritus Fellowship. This research was supported by the Leverhulme Trust, STFC
and by the University of Liverpool.

\appendix

\section{Details of calculations}
As mentioned in the main text, the expressions for the divergent contributions for diagrams with up to three internal lines emerging from the $T_Q$ vertex can be obtained from Ref.~\cite{fei1}. We denote by $c_a$ the simple pole coefficient from the diagram (a) in Fig. 8 of Ref.~\cite{fei1}, and so on. Once again we emphasise that by $c_a$ we mean precisely the simple pole coefficient from the Feynman integral, whereas the pole terms given in Figs.~7-9 of Ref.~\cite{fei1} also include a symmetry factor. We list here the values of $c_a$--$c_u$ as derived from Ref.~\cite{fei1}, suppressing a factor of $(16\pi^2)^{-L}$ at $L$ loops.
\begin{align}
c_a=&-1,\nn
c_b=\frac18,\quad c_c=&-\frac{7}{72},\quad c_d=\frac12,\nn 
c_e=-\frac18, \quad c_f=\frac{1}{24},\quad c_g=&\frac{5}{48},\quad c_h=c_i-\frac{47}{1296},\quad c_j=\frac{23}{432},\nn
c_k=\frac{5}{81},\quad c_l=\frac{11}{324},\quad c_m=&-\frac{19}{486},\quad c_n=c_o=\frac{11}{1296},\nn
c_p=\frac{11}{216},\quad c_q=-\frac{1}{24},\quad c_r=&\frac13\left(\zeta_3-\frac{23}{24}\right),\quad c_s=\frac13\left(\zeta_3-\frac{29}{24}\right),\quad
c_t=-\frac13,\nn
c_u=&-\frac13\left(\zeta_3-\frac13\right).
\label{simpa}
\end{align}

As we said earlier, we have found it convenient to reconstruct from Ref.~\cite{fei1} the expression for the anomalous dimension $\gamma_{\phi}$ in terms of the simple-pole coefficients of two-point diagrams. We easily find 
\begin{align}
\gamma_{\phi}=&-\frac12c_Ag^2-\frac12Nc_Cg^4-(c_C+c_B)g^4-c_Bg^3h-\frac12c_Cg^2h^2\nn
&-\frac32(c_Da_D+\ldots +c_La_L)+\ldots
\label{gamgen}
\end{align}
where (as defined earlier) $c_A$ is the simple pole coefficient from the diagram (a) in Fig.~7 of Ref.~\cite{fei1}, and so on, and furthermore the corresponding combinations of $g$ and $h$ are given by
\begin{align}
a_D=&(N+2)g^6+3g^5h+g^4h^2+g^3h^3,\nn
a_E=&g^6+2g^5h+g^4h^2,\nn
a_F=&g^6+\frac12Ng^5h+\frac12g^3h^3,\nn
a_G=&(N+2)(g^6+g^5h)+g^4h^2+g^3h^3,\nn
a_H=&\left(\frac12N+1\right)g^6+(N+1)g^5h+\frac12g^2h^4,\nn
a_I=&g^6+g^4h^2,\nn
a_J=&\frac12Ng^6+\frac12g^4h^2,\nn
a_K=&\left(\frac32N+1\right)g^6+\frac12(N+1)g^4h^2+\frac12g^2h^4,\nn
a_L=&\left(\frac14N^2+1\right)g^6+\frac12Ng^4h^2+\frac14g^2h^4.
\label{aA}
\end{align}
The values of the coefficients $c_A$ etc are given by (once again suppressing factors of $16\pi^2$)
\begin{align}
c_A=&-\frac13,\nn
c_B=-\frac19,&\quad c_C=\frac{11}{216},\nn
c_D=-\frac{7}{1296},\quad c_E=-\frac{71}{1296},\quad c_F=&\frac{103}{3888},\quad c_G=\frac{1}{81},\quad c_H=\frac{121}{3888},\nn
c_I=\frac19\left(\zeta_3-\frac{7}{36}\right),\quad c_J=-\frac{23}{11664},\quad c_K=&-\frac{103}{5832},\quad c_L=\frac{13}{11664}.
\label{simpA}
\end{align}
Inserting the values from Eqs.~\eqref{aA} and \eqref{simpA} into Eq.~\eqref{gamgen}, and also specialising to the fixed-point couplings in Eq.~\eqref{fixval}, we easily find the results given in Ref.~\cite{fei1}
\begin{align}
\gamma_{\phi}=&\frac{1}{N}\left[\epsilon-\frac{11}{12}\epsilon^2-\frac{13}{144}\epsilon^3\right]+\frac{1}{N^2}\left[44\epsilon-\frac{835}{6}\epsilon^2+\frac{6865}{72}\epsilon^3\right]\nn
&+\frac{1}{N^3}\left[1936\epsilon-16352\epsilon^2+\frac12(54367-7344\zeta_3)\epsilon^3\right]+\ldots
\end{align}
Using Eq.~\eqref{gamdef}, the $1/N$ and $1/N^2$ terms in the expansion may be seen\cite{fei1} to agree with those for $\eta_1$ and $\eta_2$ in Eq.~\eqref{epsex}. Furthermore, as mentioned earlier, in Ref.~\cite{fei1} the $1/N^3$ terms were shown to agree with the result for $\eta_3$ in Ref.~\cite{vas3}.

\begin{figure}
\centering
\includegraphics[width=\columnwidth]{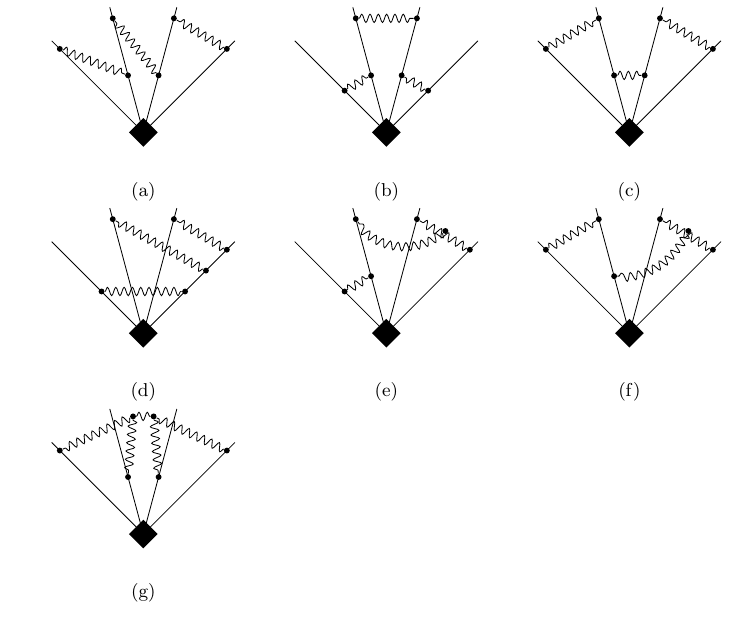}
\caption{Three-loop diagrams at leading $Q$ for $\gamma_{T_{Q}}$}\label{diagtwo}
\end{figure}

The one-loop result for the scaling dimension of $T_Q$ was given in terms of simple pole coefficients in Eq.~\eqref{Del1}. The corresponding two-loop result in terms of the simple pole coefficients is
\begin{align}
\Delta_Q^{(2)}=&-\frac13Q(Q-1)(Q-2)(6c_bg^4+c_dg^3h)\nn
&-Q(Q-1)\Bigl[\frac12Nc_cg^4+(3c_b+2c_c+c_d)g^4+2c_bg^3h+\frac12c_cg^2h^2\Bigr]\nn
&-Q\left[\frac12Nc_Cg^4+(c_C+c_B)g^4+c_Bg^3h+\frac12c_Cg^2h^2\right].
\label{Del2}
\end{align}
At three loops, the divergences for diagrams with two or three internal lines emerging from the $T_Q$ vertex can be extracted from Ref.~\cite{fei1}, as described earlier. However, at three loops there are diagrams with four internal lines which give the contribution leading in $Q$. These are depicted in Fig.~\ref{diagtwo}. We emphasise that vertices are always denoted by a small circle; a crossing of two propagators without such a circle, as seen in Figs.~\eqref{diagtwo}(d)--(g), is not a vertex.
Three-loop diagrams with four internal lines of course did not form part of the $\beta$-function computation and therefore no diagrams of this structure were evaluated in Ref. ~\cite{fei1}. We have therefore been obliged to compute their divergences {\it ab initio}. The procedure was largely straightforward. Since they are all only logarithmically divergent, we may use ``infra-red rearrangement'' which entails judiciously setting external momenta to zero, leaving only one incoming and one outgoing momentum. This should if possible be done in such a way as to avoid introducing spurious infra-red divergences. Ultraviolet subdivergences are subtracted using the $\bar{R}$ procedure. In the case of Fig.~\ref{diagtwo}(b) it was not possible to avoid spurious infra-red divergences, and we augmented the process with the $\bar{R}^*$ procedure of using a modified infra-red convergent propagator. All these methods are comprehensively described in Ref.~\cite{klein} and also summarised in (for instance) Ref.~\cite{JJ}. We have denoted the simple pole coefficients corresponding to the diagrams in Fig.~\ref{diagtwo}(a)-(g) by $d_{a-g}$ in order to avoid confusion with the coefficients extracted from Ref.~\cite{fei1}. These new coefficients are given by
\be
d_a=-\frac18,\quad d_b=\frac{1}{16},\quad d_c=\frac{5}{48},\quad d_d=d_e=-\frac{1}{24},\quad d_f=-\frac18,\quad d_g=-\frac13.
\label{dcoeff}
\ee
The final result for the three-loop contribution to the anomalous dimension of $T_Q$ in terms of simple pole coefficients is 
\begin{align}
\Delta_Q^{(3)}=&3Q(Q-1)(Q-2)(Q-3)\Bigl[\left(d_a+\frac12d_b+\frac12d_c+d_d\right)g^6\nn
&+\frac12(d_e+d_f)g^5h+\frac18d_gg^4h^2\Bigr]\nn
&+3Q(Q-1)(Q-2)(c_ea_e^{(3)}+\ldots +c_ua_u^{(3)})\nn
&+\frac32Q(Q-1)(c_ea_e^{(2)}+\ldots+ c_ua_u^{(2)})\nn
&-\frac32Q(c_Da_D+\ldots +c_La_L),
\label{Del3}
\end{align}
where the combinations of $g$ and $h$ for diagrams with two internal lines emerging from the $T_Q$ vertex are denoted as $a_e^{(2)}$ etc and given by
\begin{align}
a_e^{(2)}=&3g^6+4g^5h+2g^4h^2,\nn
a_f^{(2)}=&2(N+3)g^6+8g^5h+2g^4h^2+2g^3h^3,\nn
a_g^{(2)}=&3g^6+4g^5h+g^4h^2,\nn
a_h^{(2)}=&\left(\frac12N+2\right)g^6+2g^5h+\frac12g^4h^2,\nn
a_i^{(2)}=&(N+4)g^6+(N+2)g^5h+g^4h^2+g^3h^3,\nn
a_j^{(2)}=&\left(\frac12N+2\right)g^6+(N+2)g^5h+\frac12g^2h^4,\nn
a_k^{(2)}=&(N+4)g^6+(N+2)g^5h+g^4h^2+g^3h^3,\nn
a_l^{(2)}=&\frac12(N+4)g^6+Ng^5h+\frac12g^4h^2+g^3h^3,\nn
a_m^{(2)}=&2(N+1)g^6+\frac12(N+2)g^4h^2+\frac12g^2h^4,\nn
a_n^{(2)}=&\left(\frac14N^2+2\right)g^6+\frac12Ng^4h^2+\frac14g^2h^4,\nn
a_o^{(2)}=&(N+1)g^6+g^4h^2,\nn
a_p^{(2)}=&(N+4)g^6+g^4h^2,\nn
a_q^{(2)}=&3g^6+2g^5h,\nn
a_r^{(2)}=&2g^6+2g^5h,\nn
a_s^{(2)}=&3g^6+2g^4h^2,\nn
a_t^{(2)}=&2g^6+g^4h^2,\nn
a_u^{(2)}=&5g^6+4g^5h+g^4h^2.
\label{a2}
\end{align}
and for those with three internal lines emerging from the $T_Q$ vertex are denoted as $a_e^{(3)}$ etc and given by
\begin{align}
a_e^{(3)}=&2g^6+g^5h,\nn
a_f^{(3)}=&4g^6+g^5h,\nn
a_g^{(3)}=&2g^6+2g^5h,\nn
a_h^{(3)}=&\frac12Ng^6+\frac12g^4h^2,\nn
a_i^{(3)}=&2g^6,\nn
a_j^{(3)}=&0,\nn
a_k^{(3)}=&\left(\frac12N+1\right)g^6+\frac12g^4h^2,\nn
a_l^{(3)}=&g^6,\nn
a_m^{(3)}=&0,\nn
a_n^{(3)}=&0,\nn
a_o^{(3)}=&0,\nn
a_p^{(3)}=&\left(\frac14N+\frac12\right)g^5h+\frac14g^3h^3,\nn
a_q^{(3)}=&g^6+\frac12g^5h+\frac12g^4h^2,\nn
a_r^{(3)}=&\frac16Ng^6+\frac12g^5h+\frac16g^3h^3,\nn
a_s^{(3)}=&g^6+\frac12g^5h,\nn
a_t^{(3)}=&\frac16g^6+g^5h,\nn
a_u^{(3)}=&2g^6+g^5h+g^4h^2.
\label{a3}
\end{align}
As explained in the main text, upon inserting the values of the simple pole coefficients from Eqs.~\eqref{simpa}, \eqref{simpA}, \eqref{dcoeff}, the expressions Eqs.~\eqref{Del2} and \eqref{Del3} lead to Eqs.~\eqref{Dela} and \eqref{Delb} respectively.

\end{document}